\documentclass[prodmode, acmec]{ec12-acmsmall} 

\usepackage[ruled]{algorithm2e}

\SetAlFnt{\small}
\SetAlCapFnt{\small}
\SetAlCapNameFnt{\small}
\SetAlCapHSkip{0pt}
\IncMargin{-\parindent}

\usepackage{graphicx}
\usepackage{url}

\acmYear{2012}
\acmMonth{06}

\usepackage[caption=false]{subfig} 

\usepackage{color}

\newcommand{\sumdot}{\text{\tiny$\bullet$}}
\newcommand{\cdott}{\hspace{-2pt} \cdot \hspace{-2pt}}

\begin{document}

\markboth{E. Bakshy et al.}{Social Influence in Social Advertising}

\title{Social Influence in Social Advertising: \\Evidence from Field Experiments}
\author{EYTAN BAKSHY$^\dag$
\affil{Facebook}
DEAN ECKLES$^\dag$
\affil{Stanford University \& Facebook}
RONG YAN
\affil{Facebook}
ITAMAR ROSENN
\affil{Facebook}
}

\begin{abstract}
Social advertising uses information about consumers' peers,
including peer affiliations with a brand, product, organization, etc.,
to target ads and contextualize their display.
This approach can increase ad efficacy for two
main reasons: peers' affiliations reflect unobserved consumer
characteristics, which are correlated along the social network;
and the inclusion of social cues (i.e., peers' association with a brand) alongside
ads affect responses via social influence processes.
For these reasons, responses may be increased when multiple
social signals are presented with ads, and when ads are affiliated with
peers who are strong, rather than weak, ties.

We conduct two very large field experiments that identify the effect
of social cues on consumer responses to ads, measured in terms of ad clicks
and the formation of connections with the advertised entity.
In the first experiment, we randomize the number of social cues
present in word-of-mouth advertising, and measure how responses
increase as a function of the number of cues. 
The second experiment examines the effect of augmenting traditional ad units with a
minimal social cue (i.e., displaying a peer's affiliation below an ad in light grey
text).  On average, this cue causes significant increases in ad performance.
Using a measurement of tie strength based on the total
amount of communication between subjects and their peers, we show that
these influence effects are greatest for strong ties.
Our work has implications for ad optimization, user interface design, and
central questions in social science research.
\end{abstract}



\begin{bottomstuff}
$\dag$ Authors contributed equally to this work.\\Correspondence may be sent to E.B. at ebakshy@fb.com or D.E. at deaneckles@fb.com.\\
\end{bottomstuff}

\maketitle

\section{Introduction}

Social media activity now constitutes a substantial fraction of
time spent on the Web~\cite{goel2012browsing}.
Users of social networking technologies create explicit
representations of their relationships with other users 
(their peers)~\cite{ellison2007social}, and use those connections as
channels for information dissemination.
They also establish connections with other entities in order
to express their identities and subscribe to content~\cite{sun_09}.
The widespread adoption of such technologies has led to advertising approaches that
differ from existing approaches, such as search-based advertising.
For example, rather than inferring consumer intent
via search terms, social advertising systems can match ads to consumers
who have peers that are affiliated with the brand, product, or
organization being 
advertised~\cite{hill2006,tucker2012socialads}.
\looseness -1

Social advertising systems can also display social context about
peers who are affiliated with the entity being advertised.
These \emph{social cues} create a channel for consumers to influence
one another. Like word-of-mouth (WOM) and ``viral'' marketing approaches,
advertisers employ social ads with the goal of spreading
attitudes and behaviors through consumers' social networks.
Thus, many of the central research questions in WOM research
apply to social advertising, and research on social advertising
can contribute to the broader understanding of social influence in the
behavioral and economic sciences.
In particular, studies of social advertising can be used to learn
about how consumer responses depend on
(a) the number of social signals that consumers receive from their peers
(b) the characteristics of the relationship between the consumer and their peers. 
The present research addresses both of these topics.

Despite the similarities between WOM marketing and social advertising,
there are a number of important qualities that distinguish the two.
First, a minimal, ``lightweight'' consumer behavior
(e.g., creating a connection with an entity)
is sufficient to make that consumer a source of peer influence.
Second, social influence via these systems is passive and
automatically targeted
(cf. \cite{aral2011creating}).
That is, once an individual creates
a persistent connection with an advertised entity,
social influence can occur continually without additional actions,
such as sharing messages with others.  Finally, social advertising allows
stakeholders to play a more active role in creating and funding
ad campaigns with specific text, images, or video.

We regard \emph{social advertising} as any advertising method
that uses information about consumers' social networks 
to target ads and/or provide personalized social signals.
Thus, there are two ways in which the use of social information
in advertising can affect consumer responses:
social networks encode unobserved consumer
characteristics, which allow advertisers to target likely adopters;
and the inclusion of social cues creates a new channel for
social influence.
Recent work on social advertising (e.g., \cite{tucker2012socialads})
has recognized these mechanisms, but has been largely unable
to identify the the extent in which social influence actually plays a role.
As far as we know, the present research is the first to identify
the effect of social signals from peers on consumer responses to
advertising.
We use field experiments to make this identification possible.

\subsection{Overview}
We investigate the effects of social cues in online advertising
using two very large field experiments
that randomize the number of social cues present in ad units on Facebook.
Experiment 1 examines the marginal effect of referring to additional peers
in word-of-mouth-type ad units.
We identify the average dose--response function
(more specifically, the cue--response function), for one to
three peers, whose shape and slope differs substantially from na\"{\i}ve
estimates derived from observational data.
We find that, consistent with the expectation that
social cues are a channel for positive peer effects,
showing more peers affiliated with the advertised entity can
increase positive consumer responses.

Having established the presence of substantial social influence
effects, we extend this analysis to minimal social cues about a single peer.
Experiment 2 manipulates whether ads include a reference to a single peer
or a visually commensurate control message. Even this minimal cue ---
mentioning a single peer --- can
cause substantial increases in clicks and
the creation of connections with the advertised entity.
Finally, we examine how consumer responses vary with the strength of
the relationship between the consumer and affiliated peer, as measured by
past communication behavior. Our results show that those with stronger
ties who are affiliated with the advertised entity are more likely to
respond to an ad, even if no cue is given.  Furthermore, the presence
of the cue for strong ties can generate larger influence effects compared
to those with weak ties.

Before reporting on these experiments, we describe the
causal relationships that we aim to distinguish and estimate, and
we review some additional related work on online advertising and
social influence.

\section{Causal Relationships in Social Advertising}\label{sec:causality}
At a basic level, an advertisement is a stimulus designed to encourage
an individual to engage with an entity, such
as a brand, product, movie, musical artist, organization, etc.
The efficacy of an advertisement, typically given in terms of
a click-through rates or conversion rates, can depend critically
upon the relationship between the entity and the characteristics
of the consumer, which may include both stable traits and transient
states (e.g., demographic characteristics and recent user activity).

\begin{figure}[t]
\begin{center}
\centering
\subfloat[]{
  \label{fig:i_only}
  \includegraphics[width=0.155\textwidth]{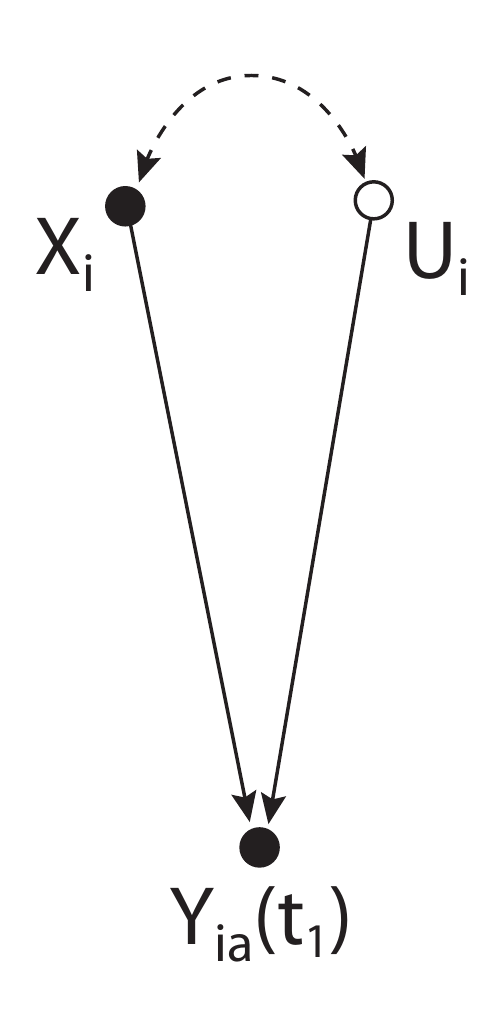}
}
\subfloat[]{
  \label{fig:homophily_only}
  \includegraphics[width=0.36\textwidth]{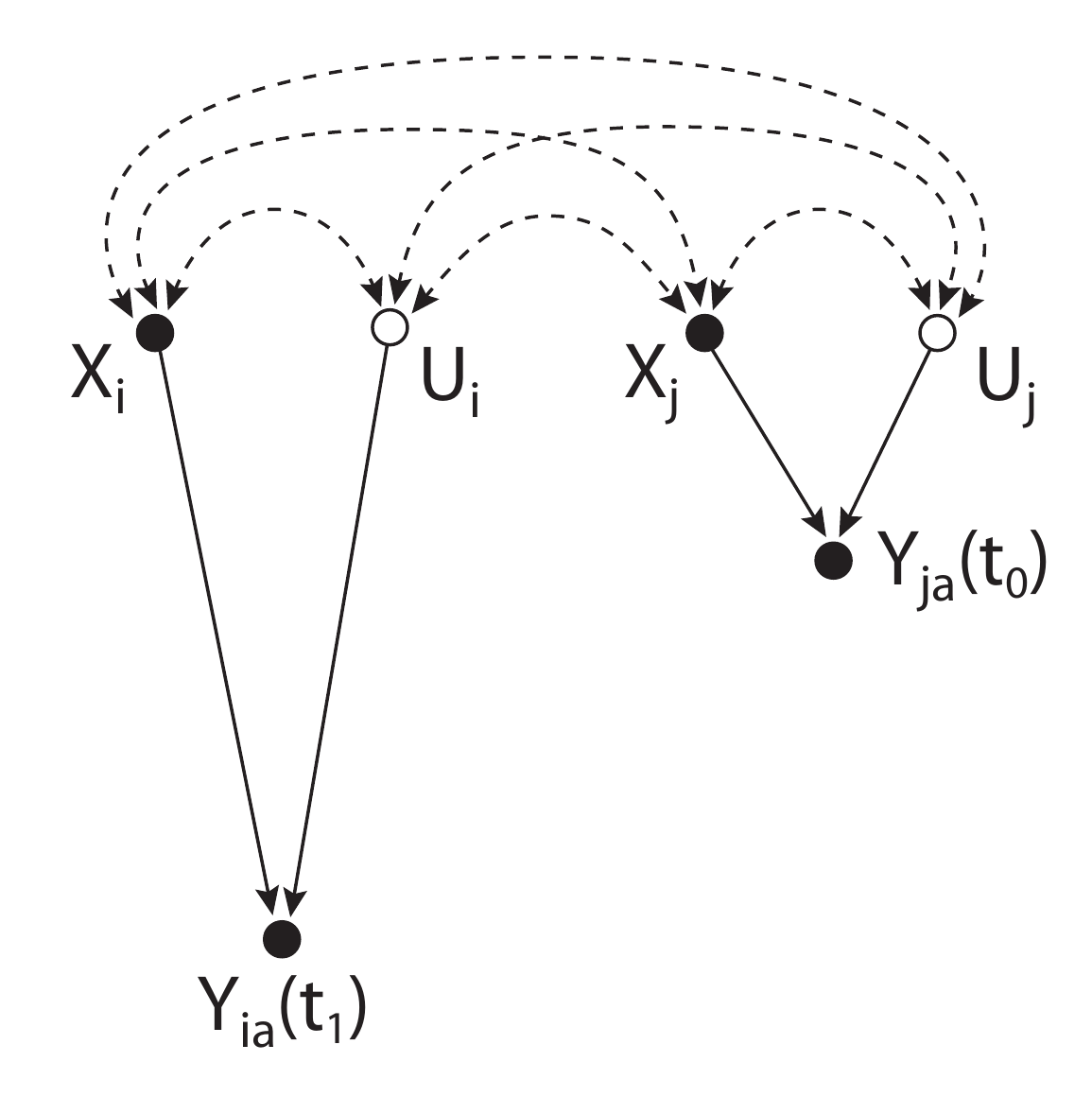}
}
\subfloat[]{
  \label{fig:full_model}
  \includegraphics[width=0.36\textwidth]{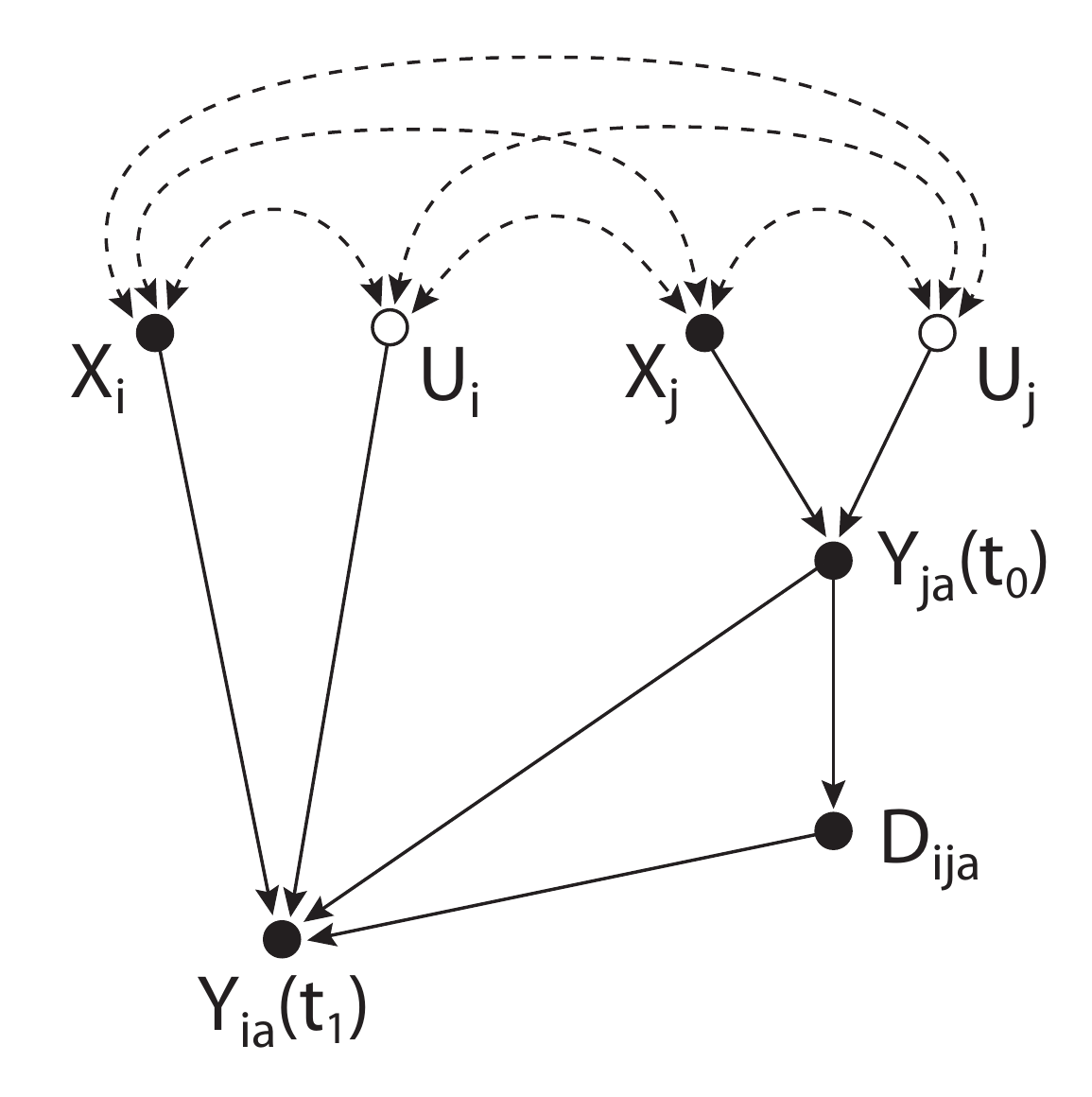}
}
\caption{Causal relationships in consumer responses to advertising.
Solid lines indicate
  cause-and-effect relationships. Dashed lines indicate
  that variables are correlated in some
  (possibly unknown) way.
(\emph{a}) Responses are caused by observed and unobserved
individual characteristics. 
(\emph{b}) Responses may be correlated with peers'
responses even when there is no social influence.
(\emph{c}) Responses can be explained both by social
influence and correlation among peer characteristics. Here one
mechanism for social influence, among other possible mechanisms,
is the inclusion of social cues, $D_{ija}$, along an ad.}
\label{fig:causal_models}
\end{center}
\end{figure}

The causal model\footnote{
We present causal models in this section by drawing 
graphs consistent with the causal
framework in \citeN{pearl_causality:_2009}.
Our discussion of confounding of homophily and influence follows
a more general treatment by \citeN{shalizi2011}.
} of this scenario (illustrated in Figure~\ref{fig:i_only})
is relatively simple:
a user $i$ has some set of known features $X_i$, which affect
the response\footnote{
$Y_{ia}$ is the response to a specific ad $a$, including the ad creative.
This allows, for the sake of simplicity, not including variables
related to the ad creative or advertised entity.}  $Y_{ia}$ to the
advertisement $a$.
Errors in predicting the response can be regarded as resulting from 
 unobserved characteristics $U_i$ of the user.

\subsection{Predictive aspects of social information}
Many characteristics, attitudes, and behaviors are clustered in 
social networks~\cite{mcpherson2001}; that is, ``birds of a feather flock together''.
This clustering has multiple causes, including
individuals preferring to associate with similar others (\emph{preference homophily}),
opportunities for forming and maintaining relationships that are
biased towards similar others (\emph{structural homophily})
\cite{kossinets09homophily,currarini_economic_2009,currarini_identifying_2010}, 
external causes whose effects are localized in the social network,
and prior peer influence by which peers become more
similar over time \cite{lewis_social_2011}.
For simplicity, we subsequently use \emph{homophily}
to refer to all observed prior clustering of consumer features.

The presence of homophily in social networks suggests that
characteristics of an individual can be predicted via characteristics
of her peers.\footnote{
Note that even if observed characteristics are incomplete or noisy,
peer characteristics can still be predictive. For example,
\citeN{backstrom_find_2010} improve estimates
of the location of individuals by using the location of their Facebook friends.}
Figure \ref{fig:homophily_only} illustrates how a peer's response to an
ad is predictive of a consumer's response, even if
information about one peer's behavior is not observed by 
and cannot affect the consumer (i.e., in the absence of peer influence). 
Here $X_i$ and $X_j$ are observed characteristics of the consumer and her
peer, and  $U_i$ and $U_j$ are unobserved characteristics.
Homophily implies that these variables will be
correlated in some way. 
Therefore, a peer $j$'s behavior at a previous time $Y_{ja}(t_0)$ is
informative about the consumer $i$'s subsequent behavior $Y_{ia}(t_1)$.
Furthermore, individuals who interact frequently exhibit greater
correlation among characteristics compared to those who do
not~\cite{mcpherson2001}.
This suggests that those who have more opportunity to influence one
another may also be expected to respond to an ad in the same way, even
in the absence of social influence.

\subsection{Peer influence and confounding}
Peer behaviors can also influence consumer responses to ads.
At least since early psychological experiments
on conformity to group behavior
\cite{sherif_psychology_1936,asch_studies_1956}
and observational studies of opinion leadership in
mass communication~\cite{katz_55}, 
these \emph{peer effects} have been a central subject
for the social and economic sciences.\footnote{
Economists sometimes refer to peer effects as
``social interactions''
\cite{durlauf_social_2010,manski_economic_2000,moffitt_policy_2001}.}

In the presence of peer effects, an individual's response to
an ad will be associated with their peers' responses not only
because of homophily, but also because peer behavior causes
individual behavior.
In particular, if consumers can observe or infer the responses
of their peers, then they are expected to use this information,
even automatically and outside of conscious awareness
\cite{tanner_chameleons_2008},
to determine their response.
Ads that include such information about peer
behaviors (i.e., as social cues) are expected
to affect consumer responses via these social influence processes.

The presence of peer effects, including those via
social cues in ad interfaces, is illustrated by 
Figure~\ref{fig:full_model}.
The peer's response to the ad $Y_{ja}(t_0)$ can affect
the consumer's response to the ad $Y_{ia}(t_1)$ via
multiple mechanisms.
Of principal interest here
is social influence via social cues,
$Y_{ja}(t_0) \rightarrow D_{ija} \rightarrow Y_{ia}(t_1)$,
although it is important to note that previous interactions between
$i$ and $j$, given by $Y_{ia}(t_0) \rightarrow Y_{ia}(t_1)$, may also cause influence effects in the absence of cues.

This picture highlights the difficulty in determining
whether a change in responses to ads is due to
the presence of social cues, other forms of social
influence stemming from prior interaction,
or the correlation in behaviors induced by homophily.
For example, even if one were to
control for all observed characteristics $X_i$ and $X_j$,
it is expected that there are unobserved characteristics
$U_i$ and $U_j$ that make the responses of
$i$ and $j$ dependent.

\subsection{Identification via experiments}
Randomized experiments are the gold standard for causal inference
\cite{rubin_estimating_1974},
and the identification of peer effects is no exception. When they are possible,
field experiments combine this internal validity with external validity 
\cite{shadish_renaissance_2009}.  Conventional experimental
methodology for online advertising
randomly assigns users to a treatment group which sees an ad and a
control or ``holdout'' group that does not see an ad; the experimenter then compares the outcome variable of interest
(such as purchases) for these two groups~\cite{randall2008}.

In the context of measuring peer effects in social advertising, experimenters can 
randomly assign user--ad pairs to receive varying numbers of social cues.
For example, when a user views an ad which can display a social cue,
whether that social cue is actually displayed would be determined by random assignment.
In the causal model of Figure~\ref{fig:causal_models}c, this amounts to
random assignment of the values of $D_{ija}$.
Experimenters can then compare response rates of user--ad pairs
assigned to different social cue conditions.
The remainder of this paper focuses on experimental comparisons
of this kind.

\section{Related work}
Online networks are focused on sharing information, and as such, have
been studied extensively in the context of information
diffusion. Large-scale observational studies explore a variety of
diffusion-like phenomena in contexts including 
the apparent spread of links on blogs~\cite{adar_05} and Twitter~\cite{bakshy2011wsdm},
the joining of groups~\cite{backstrom06},
product recommendations~\cite{leskovec06viral}, and 
the adoption of user-contributed content in virtual economies~\cite{bakshy2009}.
Data from these studies are highly suggestive of social influence: the
probability of adopting a behavior increases with
the number of adopting peers.  However, as noted
by \citeN{Anagnostopoulos08kdd} and
\citeN{aral2009}, such studies can easily overestimate the role of
influence in online behavior because of homophily.
\citeN{shalizi2011} go further to illustrate
that statistical methods cannot adequately control for
confounding factors in observational studies without the use of very strong assumptions.

Some recent work has used field experiments to examine effects of social
signals in online advertising and similar settings.
\citeN{tucker2012socialads} estimates combined effects of
social targeting and social cues in ads and highlights the value of
distinguishing them.
Two other recent experiments are similar to the present work in that they
manipulate particular mechanisms of social influence.
\citeN{aral2011creating} randomly assign individuals to
versions of an application that included or lacked viral marketing features.
Their ``passive broadcast'' feature has similarities to social advertising
in that it is visible to peers and includes a social cue.
However, their experiment manipulated whether an individual's adoption
of the application would notify their peers, rather than whether individuals' with
adopter peers were exposed to the broadcasted message.
\citeN{bakshy2012www} randomize 
exposure to particular links shared by peers on Facebook and found that
individuals were more likely to share the same links as their
friends, even if they did not see the links on the site.
This tendency was stronger for users who had multiple sharing friends,
or a single friend who was a strong tie.
Since this experiment either completely allowed exposure
or prevented it for each individual--link pair,
it did not identify the effect of influence via social cues.

\section{Setting and Data Analysis Procedures}

We conducted two large field experiments on Facebook
during a short period in 2011.
This section introduces the relevant aspects of Facebook
and some general characteristics of the data common to
both experiments.

\subsection{Subject experience}
A primary mode of interaction on Facebook occurs through News Feed,
where users share with and consume content from their
  network of peers.  Activity along these channels are called \emph{stories}, and include short messages, links,
  photos, and other content. 
Ties between users are established by one user requesting
to become \emph{friends} on Facebook, and the other user accepting
the request.
In this paper, we treat a user's friends as constituting
their set of peers.

Users can also establish connections with other entities by \emph{liking} particular \emph{pages}, which
correspond to businesses, organizations, products, movies, musical
artists, celebrities, etc. The liking affiliation is
generally visible to the user's peers when they visit the user's
profile and in peers' News Feeds. This latter case includes
connection stories that indicate a user has liked a particular
page.
Additionally, content shared by
a page appears in News Feed for users who like that page.

Several different types of ad units may be present in the right hand column
of the site. Many ad units are associated with particular pages.
These units can be targeted toward users
with peers who like the page (i.e., social targeting)
and display these peers alongside the ad unit (i.e., social cues).
For a given user--ad pair, the \emph{affiliated peers}
are all peers who like the page and are eligible
to be mentioned  (e.g., the peers' personal settings allow
displaying this information to the user).

Experiments 1 and 2 use two different types of ad units (described in
Section~\ref{sec:exp1} and Section~\ref{sec:exp2}).
In both cases, clicking on the ad unit takes the user to the advertised page.
Clicking on a link labeled ``like'' --- 
either in the ad unit itself or on the linked page ---
creates a new connection between the user and the
page. We study effects of
social cues on two responses: clicks on the
linked content and liking the advertised page.

\subsection{Assessing variation in response rates}
The observed outcomes --- responses to ad impressions --- 
are not independent and identically distributed (IID).
Users and ads vary in their response rates, and users only occur in
combination with a limited number of ads; that is, the responses
arise from a data-generating process with unbalanced
crossed random effects of users and ads.
Statistical methods that assume IID data neglect this
dependence structure and are expected
to be anti-conservative (i.e., produce confidence intervals that are too narrow).
For example, if one observed 100,000 impressions for 10,000 users
and 1,000 ads, methods that treat 100,000 as the relevant $N$ will
generally substantially overstate confidence about response rates.
To address this issue, all statistical inference in this paper employs
a bootstrap strategy for data with this crossed structure
\cite{brennan_bootstrap_1987,owen_pigeonhole_2007}.
We now briefly describe our use of this strategy so that
readers can appropriately interpret our results and apply the method
to similar problems.

In the standard (IID) bootstrap, the analyst constructs $R$ \emph{bootstrap
replicates} by sampling $N$ observations, with replacement, from a full
dataset of size $N$.  For each of these $R$ replicates, one computes the
statistic of interest (e.g., a ratio of proportions).
A simple method for computing a 95\% confidence interval for that statistic
is to use the 2.5 and 97.5 percentiles of the resulting $R$ statistics;
this is the bootstrap percentile confidence interval.

This standard bootstrap procedure is not appropriate for dependent data,
such as in our experiment, where users and ads are dependent.
Instead, analysts should resample both users and ads independently \cite{owen_pigeonhole_2007}.
We use a variation of this strategy that is suitable for online and distributed
computation~\cite{owen_bootstrapping}.
Rather than resampling $N$ observations, the data is re-weighted
according the the following procedure.\footnote{
Software implementing the multiway bootstrap in \texttt{R} is available at \url{https://github.com/deaneckles/multiway_bootstrap}. The results in this paper were produced by similar software for Apache Hive.
}
For the $r$th replicate, each user is assigned a $\text{Poisson(1)}$ draw,
and each ad is assigned a $\text{Poisson(1)}$ draw. Each user--ad pair is
then assigned the product of the corresponding draws as its weight.
This strategy is known to be conservative
when estimating the variance of means
(i.e., it produces 95\% confidence intervals that include the true mean
more than 95\% the time).
Throughout, we report 95\% bootstrap percentile confidence intervals
using $R = 500$.

\section{Experiment 1: Influence of multiple peers}\label{sec:exp1}
Theory suggests that individuals are more likely to adopt
actions previously taken by their peers when multiple social signals are
  present~\cite{schelling1973,granovetter1978,centola2007}.
  These models are central to our
  understanding of information diffusion processes and form the basis for the
  formal analysis of viral marketing~\cite{watts98collective,kempe03maxinfluence,watts2007ina}.
  However, as explained earlier,
  homophily and other factors obscure the true shape and
  magnitude of such dose--response functions in real-world data.

\begin{figure*}[pbt]
\begin{center}
\centering
\subfloat[]{
\hspace{5pt}
\includegraphics[width=0.28\textwidth]{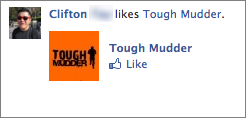} 
\hspace{5pt}
}
\subfloat[]{
\hspace{5pt}
\includegraphics[width=0.28\textwidth]{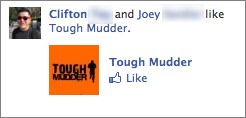}
\hspace{5pt}
}
\subfloat[]{
\hspace{5pt}
\includegraphics[width=0.28\textwidth]{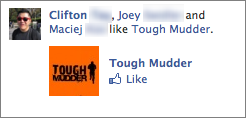}
\hspace{5pt}
}
\caption{Experimental treatment for sponsored story ad units in Experiment
  1.  Figure illustrates the three possible treatment conditions for
  users with three peers ($Z_{ia}=3$) who are
  affiliated with the sponsored page. (a) $D_{ia}=1$ (b) $D_{ia}=2$ (c)
  $D_{ia}=3$.}
\label{fig:cs_ad_unit}
\end{center}
\end{figure*}

We use \emph{sponsored story ad units} (Figure
\ref{fig:cs_ad_unit}) to understand the marginal
  effect of social signals on consumer behavior, above and beyond what
  one might expect due to homophily and other sources of
  heterogeneity.  Sponsored story ad units resemble
organic stories that appear in the News Feed
when a peer likes a page.  Similar to conventional WOM approaches, the
story does not include an advertiser-generated message, and must be
associated with at least one peer.
The main treatment unit is therefore the number of peers shown.
Since the ad units are essentially sponsored versions of organic News Feed
stories, they follow the same visual constraints imposed by the
News Feed: they must feature at least one affiliated peer,
and a small version of the first peer's profile photo\footnote{
Users select their own profile photo for the thumbnail. 
This is used throughout the site accompanying the
user's stories and comments.
}
is displayed in the leftmost part of the unit.
The first peer's name may be followed by the names of up to
two other affiliated peers.
A small thumbnail version of the page's profile image is also displayed.

\subsection{Sampling and assignment procedure}
The first experiment applies to a simple random sample of all
Facebook users. All sponsored stories displayed to users
in this sample were subject to the following assignment procedure
to determine the number of peers to be mentioned.

User--ad pairs ($i$, $a$) are randomly assigned to a number of peers $D_{ia}$ to be
mentioned in the sponsored story. Since the maximum number of peers shown is limited by the
number of affiliated peers $Z_{ia}$
(i.e., peers who have liked the advertised page and are eligible to be
mentioned), user--ad pairs are assigned with equal probability
to all of the possible values of $D_{ia} \in \{1, ..., \text{min}(Z_{ia}, 3)\}$. We limit our
analysis to user--ad pairs with one, two, or three available peers ($Z_{ia} \leq 3$).
Experimental assignment is a deterministic function of
user--ad pairs, so all impressions for a user--ad pair have the
same number and order of peers mentioned in the sponsored story. 
In total, Experiment 1 includes  23,350,087 users,  148,606 ad IDs,
and 101,633,907 distinct user-ad pairs.

\subsection{Average cue--response function}
\label{sec:exp1crf}
\begin{figure*}[pbt]
\begin{center}
\centering
\subfloat[]{\includegraphics[width=0.5\textwidth]{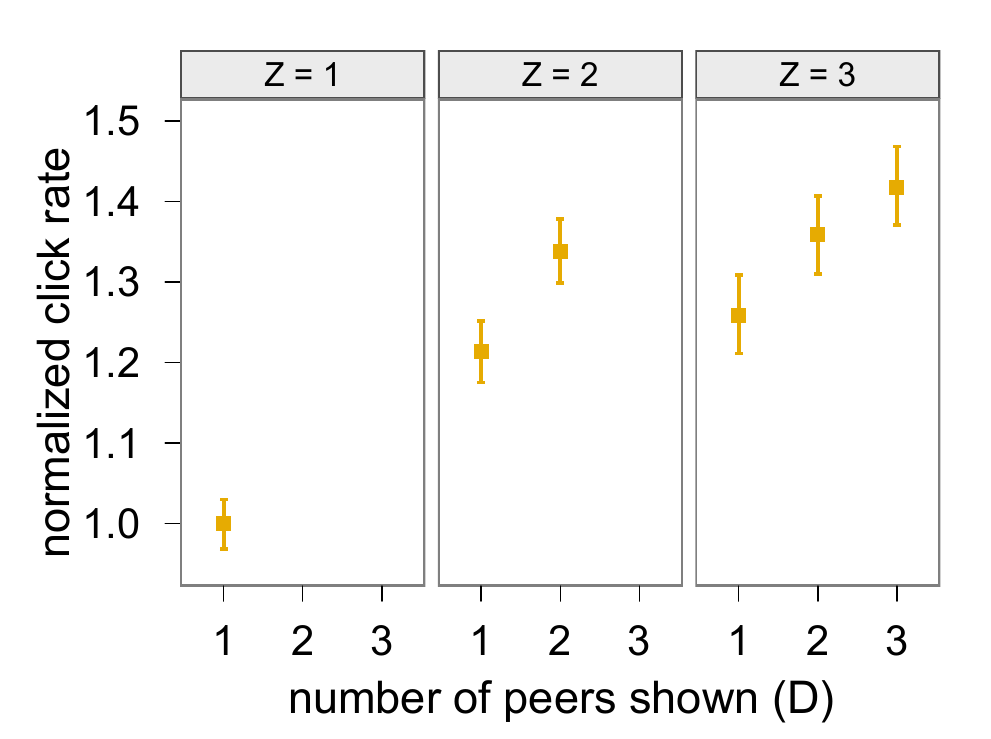}}
\subfloat[]{\includegraphics[width=0.5\textwidth]{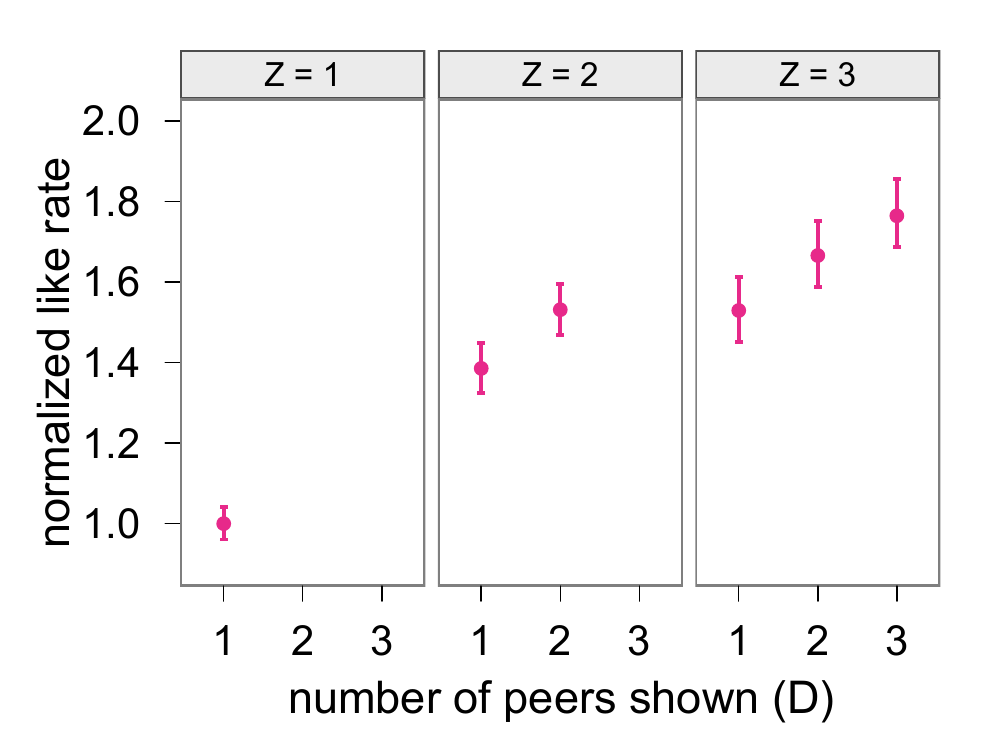}}
\caption{The average cue-response function is the relationship
  between the number
  of peers shown, $D$, and (a) click-through rate and (b) like rate.
  Within-panel differences are caused by the treatment
  (number of cues), and are significant with $p<0.005$. Error bars
  are 95\% bootstrapped confidence intervals. Note that because some
  of the same users and ad IDs contribute to
  each response rate, overlapping intervals do not indicate non-significance
  of associated differences and ratios.
  Panels
  represent different populations of user--ad pairs, where the user
  has $Z= 1, 2, {\text{or }} 3$ affiliated peers.
All probability scales are normalized by the smallest point estimate in the plot.}
\label{fig:conn_story_ctr}
\end{center}
\end{figure*}

We first examine the \emph{average cue--response} function, which
shows how response rates vary as a
function of the number of peers shown.  This relationship, conditional
on the number of affiliated peers, is shown in
Figure~\ref{fig:conn_story_ctr}.
Panels with $Z=2$ and $Z=3$ illustrate the causal effect of showing
multiple peers on click and like rates.    For user--ad pairs with two
affiliated peers, displaying a second peer caused
a 10.3\% ($CI = $ [8.7\%, 11.9\%]) relative increase in click rate and a
10.5\% ($CI = $ [8.4\%, 12.4\%]) relative increase in like rate.
When there were three affiliated peers, this increase was slightly weaker:
the click rate increased by 8.0\% ($CI = $ [5.7\%, 10.3\%)]
and like rate by 8.9\% ($CI = $ [6.0\%, 12.1\%]).

The cue--response function can also be used to examine whether
influence appears to follow simple contagion, where each additional cue is
expected to cause a constant or slightly sub--linear increase, or complex
contagion (i.e.~\cite{schelling1973,granovetter1978,centola2007}),
whereby additional social signals result in a super--linear increase
in the response.  We find that there is no statistically significant difference between the
probability increase from $D=1$ to $D=2$, and $D=2$ to $D=3$ ($p>0.1$ for clicks, $p>0.3$ for
likes).  This is consistent with simple contagion, although since we
only measure average effects, it is difficult to entirely rule out complex
contagion.

It is also apparent from Figure~\ref{fig:conn_story_ctr} that response
rates increase with the number of affiliated peers, even when users
only see a single cue.
This trend illustrates how in many
observational settings, the relationship between
an individual's behavior with the number of peers who also exhibit that behavior
may not reflect increased levels of influence.
While homophily is an obvious cause of the increase in response rates,
the difference should not be attributed to homophily alone.
For example, consider variation in the network-wide popularity
of pages: sponsored stories associated with pages with more connected users
will be observed more frequently with a larger number of affiliated peers.
If these more popular pages have sponsored stories with higher response rates,
then this will contribute to the between-panel differences in
Figure \ref{fig:conn_story_ctr}. Likewise, users with more peers
will be observed more frequently with a larger number of affiliated peers.
These users might be expected to differ in their response rates.
Finally, the selection of which stories to display to a user is endogenous: an individual's
characteristics are used to predict click-through rates, which cause
the story to be displayed, but these characteristics may also be correlated
with the number of affiliated peers.
Thus, unlike comparisons given by the average cue--response function
(within panel comparisons),
variation between users with different numbers of affiliated
peers cannot be given a straightforward interpretation.

Two limitations of this experiment concern the nature of the particular
ad unit used. First, since this unit consists of reporting the connection
of one or more peers with the entity, it was not possible to examine
the effect of social cues relative to a baseline without any cues.
Second, as the number of peers shown
increases, so does the height of the ad unit and number of non-white pixels in
the ad unit. While this change is very small relative to the overall
size of the unit, this is a potential confound with the effects of
increasing the number of peers referred to in the social cue.
Both of these limitations motivate the design of Experiment 2,
in which we examine how a more minimal social cue affects
consumer responses.

\section{Experiment 2: Influence of Minimal Social Cues}\label{sec:exp2}
The next experiment examines the presence of minimal social cues
with an ad. In particular, we measure the effect of adding light grey
text with the name of a single peer associated with the entity being advertised.
Thus, the first goal of this experiment is to identify the effect of
having a cue alongside an ad.
We test whether the
social influence effects in Experiment 1 extend to the presence or
absence of any social cues when these cues are 
visually commensurate with one another and subordinate to an advertiser-created message.

The second experiment also enables us to examine how
the effect of social cues in advertising varies with the strength of the relationship
between a consumer and affiliated peer. Previous work motivates
the hypothesis that both homophily~\cite{hill2006,kossinets09homophily}
and social influence~\cite{hovland_influence_1951,rogers_homophily-heterophily:_1970,goethals_similarity_1973}
should cause peers with stronger relationships to have
responses that are more correlated.
We expect response rates to be higher for user--ad pairs
where an affiliated peer is a strong tie rather than a weak tie.
This effect should exist even in the absence of a social cue,
but we also expect that the effect of a social cue will be larger
for strong ties than weak ties.

\begin{figure*}[]
\begin{center}
\centering
\subfloat[]{
\hspace{5pt}
\includegraphics[width=0.32\textwidth]{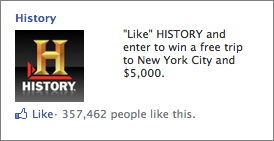}
\hspace{5pt}
}
\subfloat[]{
\hspace{5pt}
\includegraphics[width=0.32\textwidth]{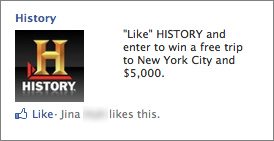}
\hspace{5pt}
}
\caption{The two treatment conditions for social ads in Experiment
  2. Subjects who are to be exposed to ads with at least one
  affiliated peer are randomly assigned to see either
(a) general information about the total number of affiliated individuals ($D_{ia} = 0$) or
(b) a minimal social cue featuring one affiliated peer ($D_{ia} = 1$).
}
\label{fig:social_ad_unit}
\end{center}
\end{figure*}

Experiment 2 manipulates social cues alongside a \emph{social ad} unit,
which includes advertiser-specified ``creative'' consisting
of a title, image, and caption (Figure~\ref{fig:social_ad_unit}).
Beneath the custom creative is a link that says ``Like'' and
gives information about people who like this page in
small gray text.
If a user has any affiliated peers, the name of one of those peers can be displayed.
If a user has no affiliated peers, the total number of users
who have liked the page is displayed.

\subsection{Sampling and assignment procedure}
The procedures for
Experiment 2 are the same as Experiment 1.
The experiment differs in that it
involves random assignment to the presence
or absence of a social cue (Figure \ref{fig:social_ad_unit}),
rather the number of peers referred to in an always-present cue.
That is, user--ad pairs $(i, a)$ are randomly assigned to the
presence of a social cue mentioning a single peer ($D_{ia} = 1$)
or to the absence of a social cue ($D_{ia} = 0$). In the latter case,
the total number of users who like the page is displayed in
the same location and typeface as the peer's name.\footnote{
The $D_{ia} = 0$ stimuli can also be regarded
 as including a social cue, though not
a personalized social cue about peers. Compared with
non-informative text in this location, we might expect that
ad efficacy will be higher with this general prevalence
information. In this case, Experiment 2 underestimates
the effects of having a minimal social cue.}  As before, the selection of
the treatment and peer is deterministic, so users assigned to the
$D_{ia}=1$ condition will always see the same peer.
In total, Experiment 2 includes 5,735,040 users, 1,155,178 ads IDs, and
137,505,771 user-ad pairs. 

\begin{figure*}[pbt]
\begin{center}
\centering
\subfloat[]{\includegraphics[width=0.51\textwidth]{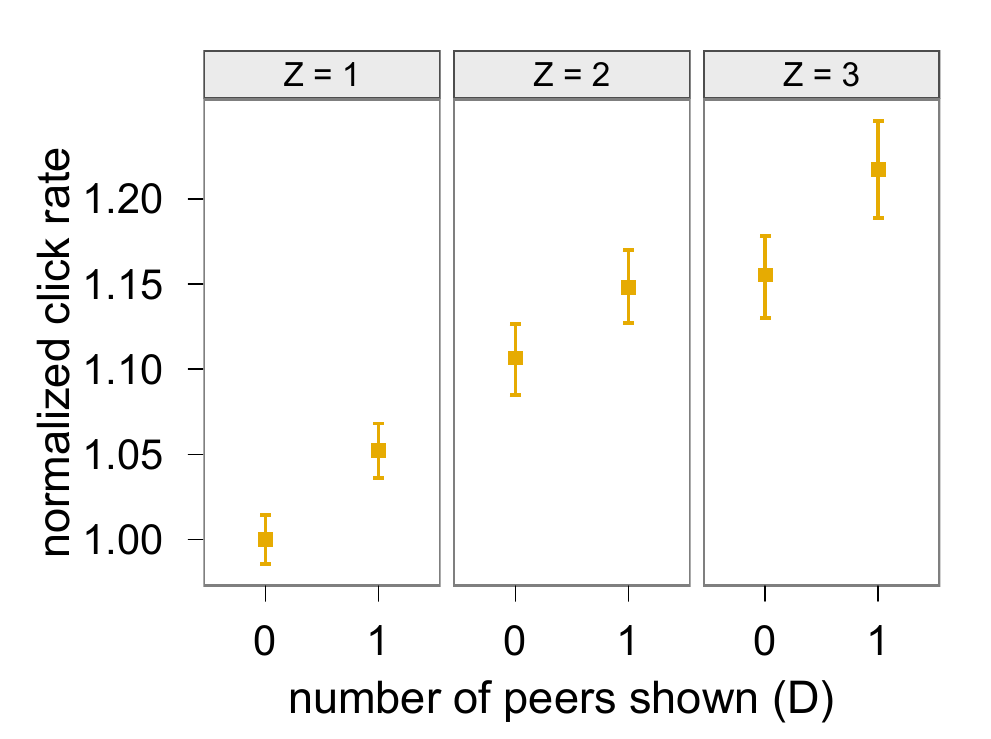}}
\subfloat[]{\includegraphics[width=0.51\textwidth]{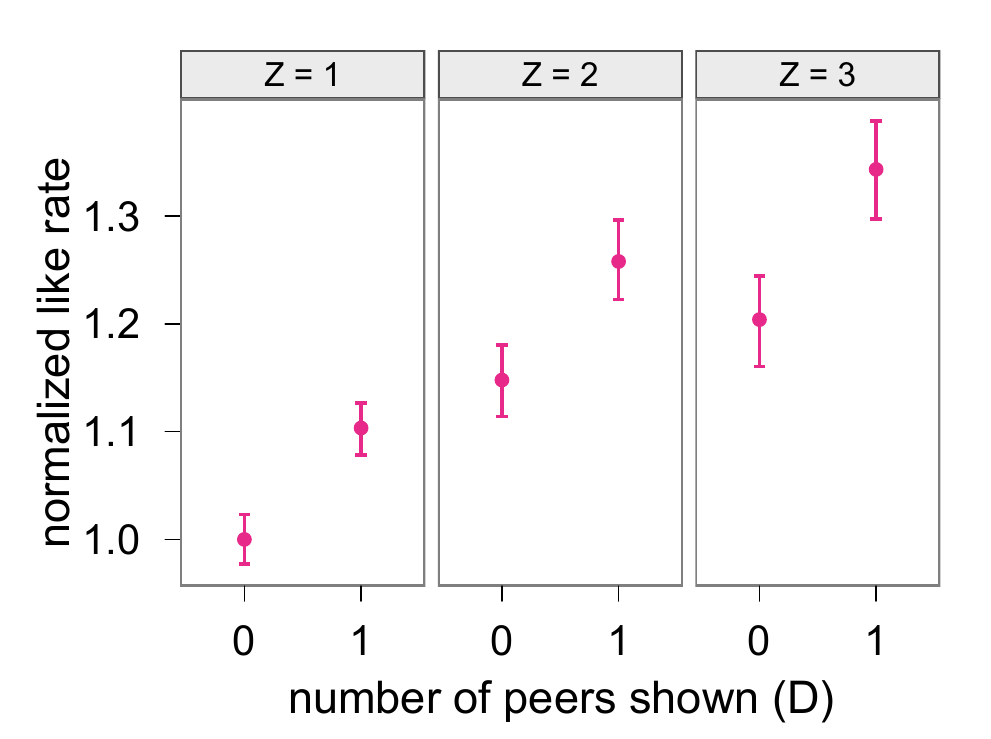}}
\caption{Average cue-response function for subjects shown no peers ($D= 0$) or one peer ($D= 1$),
when subjects have 1, 2, or 3 peers that can be associated with the advertisement.
Within panel differences reflect differences in the treatment
condition and are significant with $p<0.005$ for both click and like rates.
Error bars are 95\% bootstrapped confidence intervals.
Note that because some
  of the same users and ad IDs contribute to
  each response rate, overlapping intervals do not indicate non-significance
  of associated differences and ratios.
}
\label{fig:social_ad_ctr}
\subfloat[]{\includegraphics[width=0.48\textwidth]{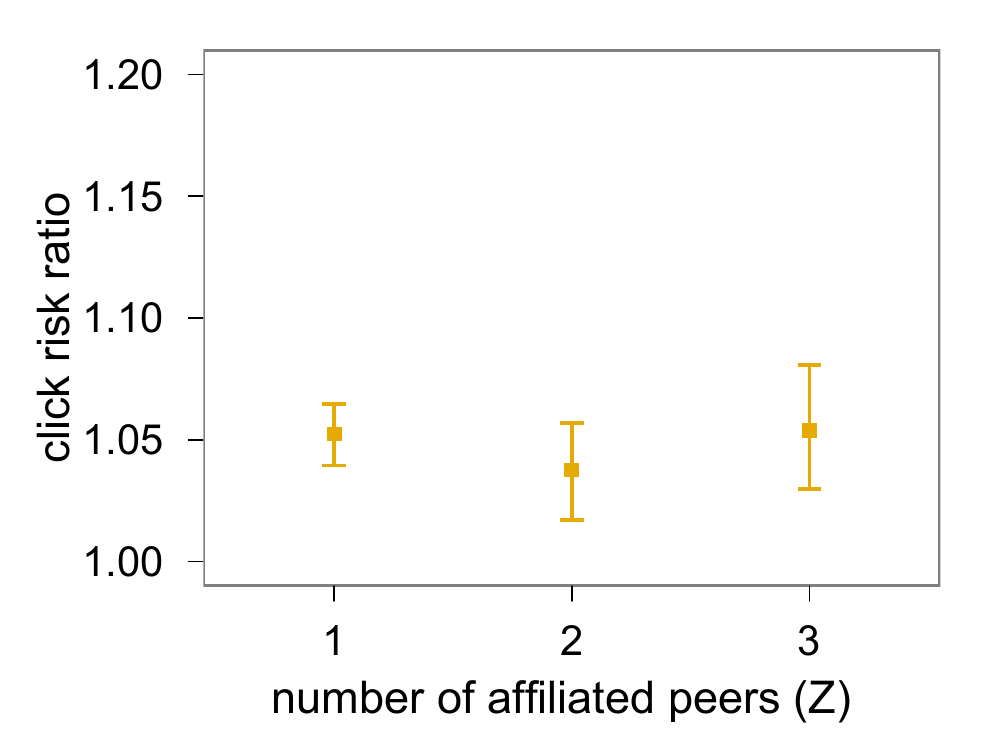}}
\subfloat[]{\includegraphics[width=0.48\textwidth]{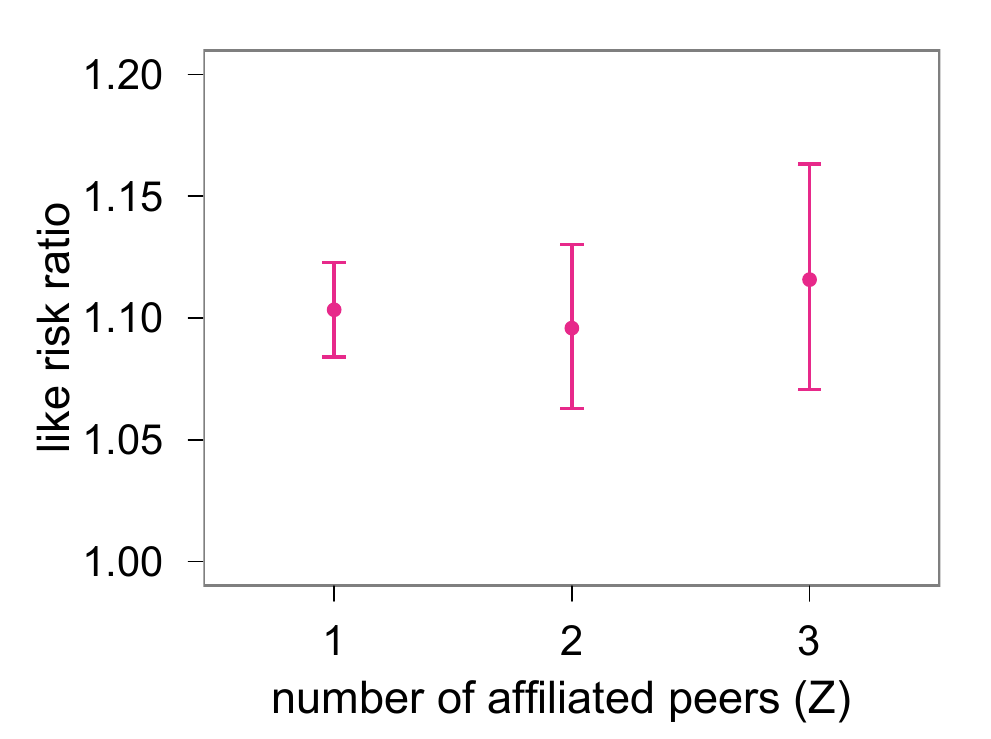}}
\caption{Relative increase in responses from
including the minimal social cue. Error bars are 95\% bootstrapped confidence intervals.}
\label{fig:social_ad_rr}
\end{center}
\end{figure*}

\subsection{Average effect of a social cue}

We begin by estimating average effects of the minimal social cue.
Figure~\ref{fig:social_ad_ctr} displays response
rates with and without the minimal social cue for different numbers
of affiliated peers. 
Similar to the previous experiment, comparisons across different numbers of
affiliated peers (i.e., between panels of Figure~\ref{fig:social_ad_ctr})
show that response rates increase when more peers are available, regardless of how many were shown.
As before, it is assumed that this increase results from homophily and heterogeneity in ad-user pairs
that are associated with the number of affiliated peers available to a user for a given ad.
(see Section~\ref{sec:exp1crf} above).

Average peer effects resulting from the minimal social cue are identified 
by comparing responses
with and without the social cue
for each number of affiliated peers.
We find that, depending on the number of affiliated peers,
the cue increases 
click rates by 3.8\% to 5.4\%
and like rates by 9.6\% to 11.6\%
(Figure~\ref{fig:social_ad_rr}).
For example, for users with a single affiliated peer,
referring to that peer increases
the click rate by 5.2\% ($CI = $ [4.0\%, 6.5\%])
and the like rate by 10.3\% ($CI = $ [8.4\%, 12.3\%]).
This provides evidence that even a minimal social cue
can substantially affect a consumer's response to an ad.

\subsection{Tie strength}
To what extent do consumer responses to an ad depend on the nature of the
relationship between the user and her affiliated peer?
The preceding analysis treats relationships among consumers as binary,
where a tie is either present or absent.
We now examine the strength of a consumer's relationship with
her affiliated peers on a many-valued scale.
To do this, we consider cases in
which users have exactly one affiliated peer who was either
shown or not shown (i.e., those user--ad pairs
featured in the first panel of Figure~\ref{fig:social_ad_ctr}).

\subsubsection{Measure of tie strength}
We use a measure of tie strength based on the frequency in which two users
communicate with each other via Facebook.
The total number of comments and messages created by a user during
this 90 day period is their \emph{total communication count} $C_{i\sumdot}$.
We measure the strength of the directed tie between user $i$ and user $j$ as
$W_{ij} = C_{ij} / C_{i\sumdot}$. We define \emph{tie strength} as the fraction of
user $i$'s communications that are directed at user $j$ or on posts by user $j$.
Similar measures are good predictors
of standard measures of interpersonal trust \cite{burke2011chi}
and selection of peers as ``top friends'' \cite{kahanda2009using}.
Recent studies of information consumption \cite{backstrom2011center}
and diffusion have also used this measure \cite{bakshy2012www}.

Because the validity of our tie strength measure requires that the user communicates
via Facebook during the 90 day period prior to the experiment, we restrict our analysis of tie strength
to users who fall in the middle of the distribution of total
communication count $C_{i\sumdot}$;
in particular, all users in this section fall within the 25th and 75th
percentiles of the count distribution.
In addition, we use a percentile-transformed total communication
count $q(C_{i\sumdot})$ as a covariate that measures 
how much user $i$ uses Facebook messages and comments to communicate.

\subsubsection{Model}

In order to pool information across similar values of tie
strength and to facilitate statistical inference,
we model responses to ads using logistic regression with natural splines.
We fit a model in which ad clicks are predicted by
the presence of a minimal social cue $D_{ia}$ for user $i$ and ad $a$,
measured tie strength $W_{ij}$ for user $i$ with the affiliated peer $j$,
and the user's percentile-transformed total communication count $q(C_{i\sumdot})$.
This model includes interactions of the minimal social cue
with tie strength and with total communication count.

In particular, the model is specified as
\begin{equation}
\label{eq:ts_model}
Y_{ija} \sim \alpha + \delta D_{ia} + \tau f(W_{ij}) + 
\eta D_{ia} \cdott f(W_{ij}) + 
\gamma q(C_{i\sumdot}) + \lambda q(C_{i\sumdot}) \cdott f(W_{ij})
\end{equation}
where $f$ is a natural spline basis expansion for measured tie strength with knots
at the second and third quartiles of measured tie strength over all impressions.
We fit the same model to the data for the response of liking the page.\footnote{
Other related model specifications yield qualitatively similar results. Specifically,
models also including a three-way product interaction of $D_{ia}$,
$q(C_{i\sumdot})$, and $f(W_{ij})$ resulted in very similar fits. Models with only
a linear term for tie strength $W_{ij}$, rather than a natural spline,
have a statistically significant interaction between tie
strength and the cue, but suffered
from model bias since the true relationship is non-linear on the logistic scale.
}
\begin{figure*}[]
\begin{center}
\centering
\subfloat[]{\includegraphics[width=0.5\textwidth]{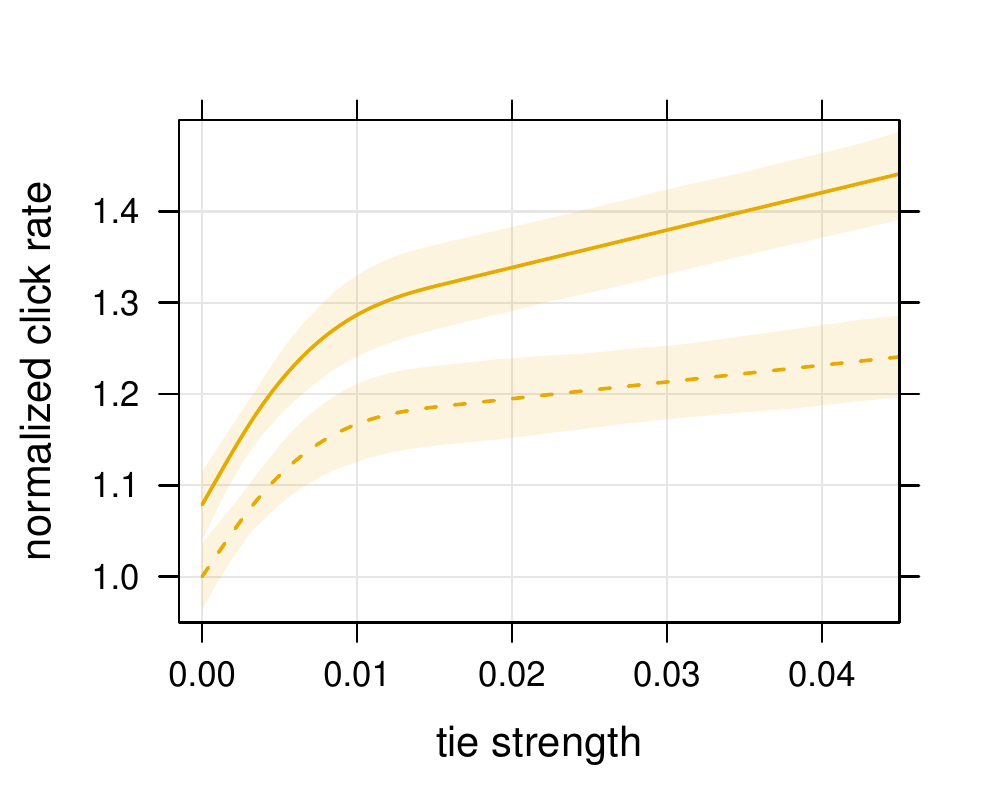}}
\subfloat[]{\includegraphics[width=0.5\textwidth]{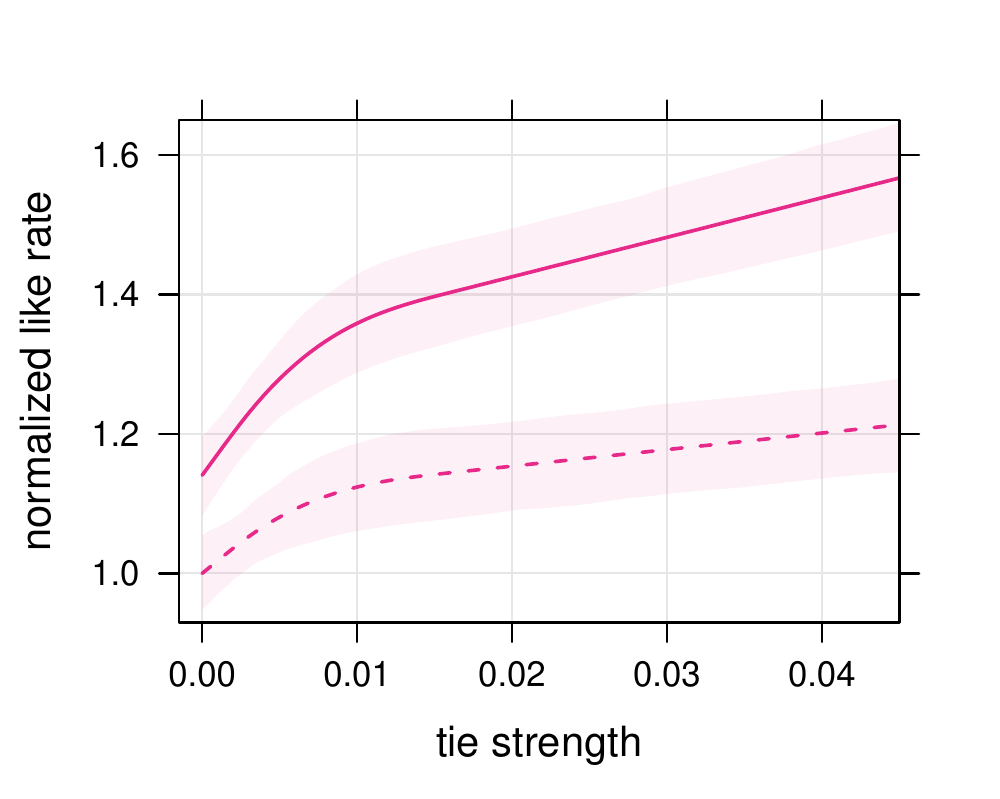}}
\caption{Estimated average response as a function of tie strength between the user
 and the single affiliated peer. Action rates increase with tie strength both in 
 the presence ($D = 1$, solid) and absence ($D = 0$, dashed)
 of the minimal social cue featuring the affiliated peer.
 Each plot shows model fits (via Equation \ref{eq:ts_model})
 for users at the median total communication count (i.e., $q(C_{i\sumdot}) = 0.5$), ranging from zero to the 90th percentile of tie strength. 
 Shaded regions are 95\% bootstrapped confidence intervals of the
 predicted response rate, which are generated by fitting the model to
 $R=500$ bootstrap replicates of the data.}
\label{fig:social_ad_ts_ctr}

\subfloat[]{\includegraphics[width=0.5\textwidth]{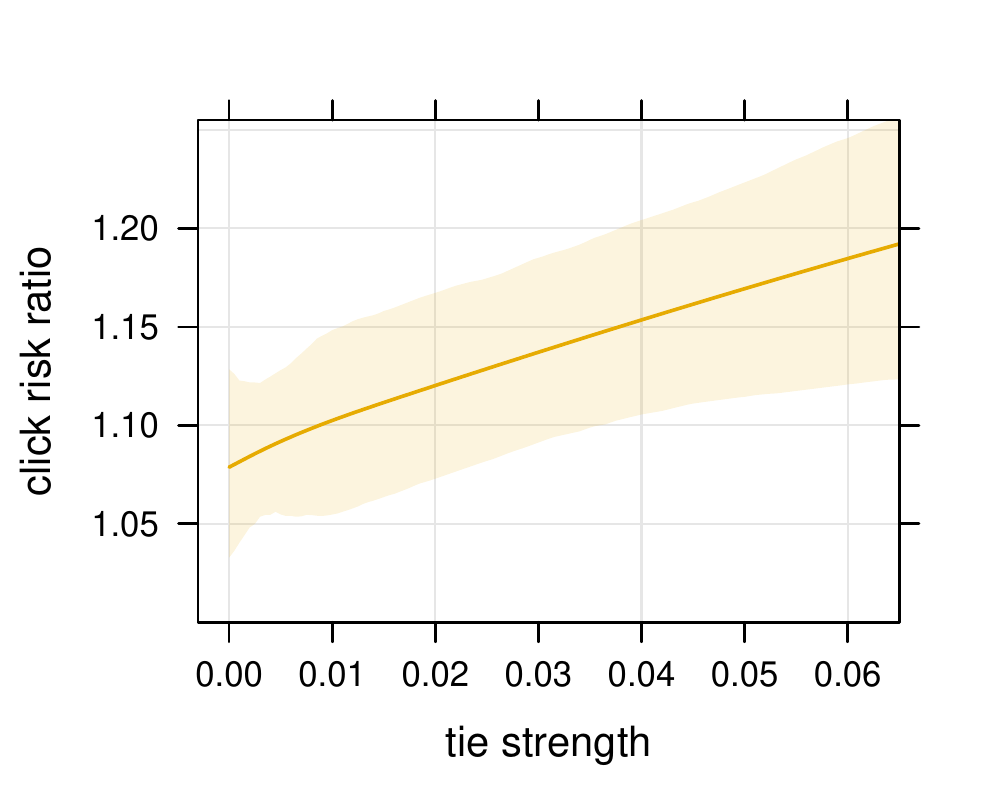}}
\subfloat[]{\includegraphics[width=0.5\textwidth]{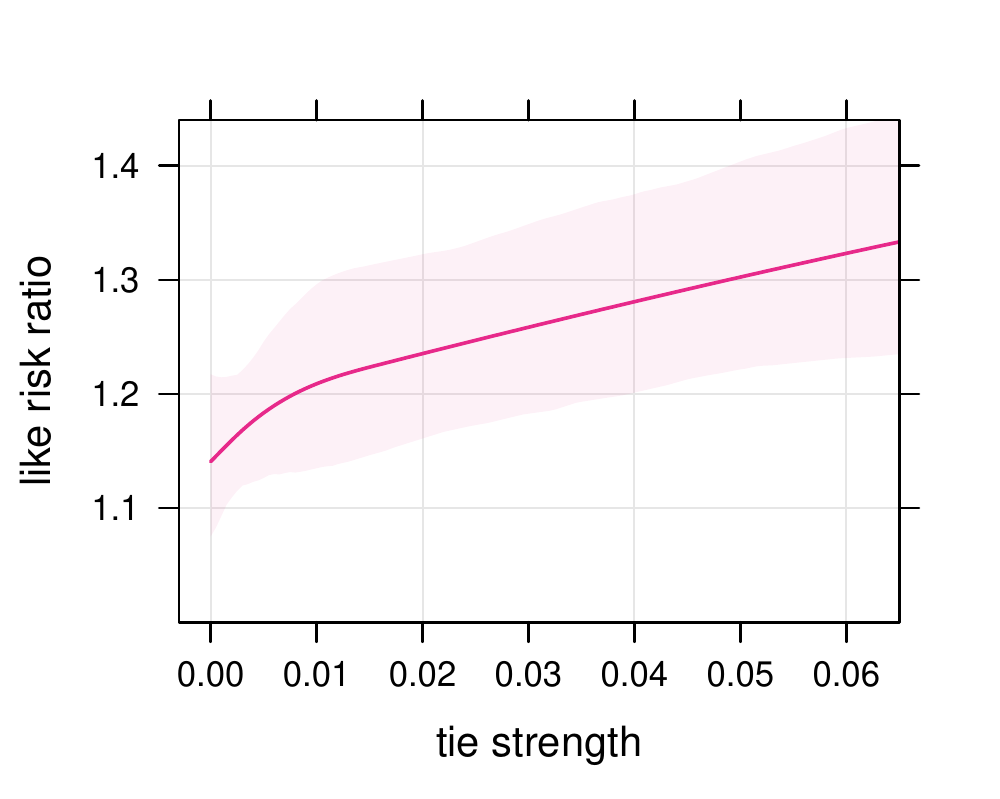}}
\caption{Estimated risk ratio of clicking 
and liking 
as a function of measured tie strength from the model in Equation \ref{eq:ts_model}.
Individuals with an affiliated peer with whom they communicate with
more often have a higher relative increase in the probability of each response.
Shaded regions are 95\% bootstrapped confidence intervals of the ratio between predicted responses with and without the cue.}
\label{fig:social_ad_ts_rr}
\end{center}
\end{figure*}

\subsubsection{Results}

Response rates increase with tie strength
both in the presence and absence of social cues. 
Figure \ref{fig:social_ad_ts_ctr} shows predicted response rates for user--ad
pairs as a function of tie strength with (solid) and without (dashed)
the minimal cue.\footnote{
For both responses, we fit the model separately to each of the $R = 500$ bootstrap replicates. Confidence intervals are computed as the 2.5 and 97.5 percentiles of the bootstrap distribution of predicted response rates.}
Since response rates vary with total communication count,
we plot the results for users at the median
of the total communication count, $C_{i\sumdot}$.
The increase of response rates in both conditions
is consistent with the expectation that social influence, homophily,
and other sources of heterogeneity should all produce increased correlation
in responses among strong tie peers compared to weak tie peers.

If influence is greater for strong ties, then we should expect the
increase in response rates that results from the display of social cues to be larger for 
strong ties than weak ties.  We find that this is the case, both in terms
of risk ratio and risk difference (i.e., average treatment effect on the treated).
Figure \ref{fig:social_ad_ts_rr} displays the click and like risk ratios
for different values of measured tie strength. These risk ratios
increase with tie strength.
For example, consider ad--user pairs with $W_{ij} = 0$ and $W_{ij} = 0.045$ (the 90th percentile). The stronger tie ad--user pairs
have a larger click risk ratio (risk ratio difference of 0.083, $CI =$ [0.014, 0.160])
and a larger like risk ratio (risk ratio difference of 0.151, $CI =$ [0.036, 0.280]).
That is, the relative increase from the social cue
is larger for stronger ties, even though the denominator
of this ratio --- the response probability without the social cue ---
is also larger for stronger ties.

\section{Conclusion}\label{sec:discussion}
We summarize the primary contributions of this work as follows.
First, we rigorously measure social influence via social cues on an
economically relevant form of user behavior.  We construct
the average cue-response function, which gives the relationship
between the number of social signals received by the individual and average
rates of response.   The shape and slope of the cue--response function
differ dramatically from the na\"{\i}ve observational estimates of social
influence effects obtained through simple conditioning on the number
of peers; this difference illustrates the utility of experimentation in estimating
peer effects.
Second, we demonstrate the substantial consequences of
including minimal social cues in advertising. This highlights
that the most subtle of forms of personalized social signals can play an important
role in determining consumer responses to advertising, above and
beyond correlations due to homophily.
Third, we measure the positive relationship between a consumer's response
and the strength of their connection with an affiliated peer.
This relationship exists even when social cues
are absent from advertisements.  Furthermore, cues
have a stronger effect for stronger ties. These results suggest that social advertising
systems can benefit from incorporating tie strength measures into the
selection of ads and social cues.
Finally, we  hope that the explicit analysis of causal
relationships in social advertising will aid researchers and
decision-makers to understand recent developments in marketing.

The present work has important limitations that
suggest directions for future research.
First, tie strength as a theoretical construct has been given
multiple definitions and is operationalized in many ways.
We used transactional data about communication behavior
as a measure of tie strength. Our choice of measure provides a readily
interpretable and applicable measure, but limited some
analyses to consumers with sufficient levels of communication
activity, and future work could attempt to replicate our results for
other measurements of tie strength, including trust and intimacy.
Second, we could not randomize the
tie strength of the peers who were affiliated with and referred
to in the social cues. In particular, the population of ad--user
pairs is different for strong and weak ties, so the analysis of
Experiment 2 did not directly allow for causal inference about
the effects of changes to tie strength of peers in social cues.
Third, we have only examined two methods for presenting social cues with online advertising.
The minimal social cue we examined in Experiment 2 
is appealing in that it  allows us to attribute
its affects to social influence processes, rather than simply
increasing the size or general visual characteristics of the ad.
However, many other ways of presenting social information
to consumers are possible.
For example, a more prominent display
of peers' association or activity around an entity, combined with an
advertiser-generated creative, may have a stronger response than any
one of the two units on their own. In addition, future work may examine the
effectiveness of such ad displays, as well the impact of social
signals on other outcomes,
such as ad recall and attitudes towards brands~\cite{gibs2010}.
Finally, it is important to note that we only estimate average
effects over the population of Facebook users that naturally see ads
with social context.  Individual responses to cues may vary
substantially from user to user, and further work is needed to understand how factors
such as age, gender, number of friends, or activity on site relate to
the effect of social cues.

\bibliographystyle{acmsmall}

\end{document}